\documentclass[aps,pra,twocolumn,showpacs,superscriptaddress]{revtex4-1}

\expandafter\let\csname equation*\endcsname\relax
\expandafter\let\csname endequation*\endcsname\relax

\usepackage{epsfig}
\usepackage{bm}
\usepackage{xfrac}
\usepackage{color}
\usepackage{graphicx}
\usepackage{tikz}
\usepackage{lipsum}
\usepackage{listings}
\usepackage{tabularx,bbding}

\pdfoutput=1
\usepackage[utf8]{inputenc}
\usepackage[english]{babel}
\usepackage[T1]{fontenc}
\usepackage{amsmath,ulem}
\usepackage{hyperref}

\usepackage{tikz}
\usepackage{lipsum}

\usepackage{pifont}
\newcommand{\cmark}{\ding{51}}

\usepackage{enumitem}
\usepackage{graphicx}
\usepackage{xcolor}
\usepackage{qcircuit}
\usepackage{braket}
\usepackage[ruled,vlined]{algorithm2e}

\usepackage{color, colortbl}
\definecolor{LightCyan}{rgb}{0.88,1,1}
\definecolor{Gray}{gray}{0.95}

\usepackage[colorinlistoftodos]{todonotes}
\definecolor{codegreen}{rgb}{0,0.6,0}
\definecolor{codegray}{rgb}{0.5,0.5,0.5}
\definecolor{codepurple}{rgb}{0.58,0,0.82}
\definecolor{backcolour}{rgb}{0.95,0.95,0.92}

\lstdefinestyle{mystyle}{
    commentstyle=\color{codegreen},
    keywordstyle=\color{magenta},
    numberstyle=\tiny\color{codegray},
    stringstyle=\color{codepurple},
    basicstyle=\ttfamily\footnotesize,
    breakatwhitespace=false,         
    breaklines=true,                 
    captionpos=b,   
    frame=lines,
    keepspaces=true,                 
    numbers=left,                    
    numbersep=5pt,                  
    showspaces=false,                
    showstringspaces=false,
    showtabs=false,                  
    tabsize=2,
    xleftmargin=1.8em
}

\AtBeginDocument{%
  \providecommand\BibTeX{{%
    \normalfont B\kern-0.5em{\scshape i\kern-0.25em b}\kern-0.8em\TeX}}}

\begin{document}

\title[Qsun: an open-source ...]{Qsun: an open-source platform towards practical quantum machine learning applications}

\author{Chuong Nguyen Quoc}
\thanks{Electronic address: quoc.chuong1413017@gmail.com}
\affiliation{Vietnamese-German University, Ho Chi Minh City,
Vietnam, 70000}

\author{ Le Bin Ho}
\thanks{Electronic address: binho@riec.tohoku.ac.jp}
\affiliation{Ho Chi Minh City Institute of Physics, 
National Institute of Mechanics and Informatics, 
Vietnam Academy of Science and Technology, Ho Chi Minh City, 700000, Vietnam}
\affiliation{Research Institute of Electrical Communication, Tohoku University, Sendai, Japan, 980-8577}

\author{Lan Nguyen Tran}
\affiliation{Ho Chi Minh City Institute of Physics, 
National Institute of Mechanics and Informatics, 
Vietnam Academy of Science and Technology, Ho Chi Minh City, 700000, Vietnam}

\author{Hung Q. Nguyen}
\thanks{Electronic address: hungngq@hus.edu.vn}
\affiliation{Nano and Energy Center, VNU University of Science, Vietnam National University, Hanoi, Vietnam, 120401}

\vspace{10pt}

\date{\today}

\begin{abstract}
Currently, quantum hardware is restrained by noises and qubit numbers. Thus, a quantum virtual machine that simulates operations of a quantum computer on classical computers is a vital tool for developing and testing quantum algorithms before deploying them on real quantum computers.
Various variational quantum algorithms have been proposed and tested on quantum virtual machines to surpass the limitations of quantum hardware.
Our goal is to exploit further the variational quantum algorithms towards practical applications of quantum machine learning using state-of-the-art quantum computers.
In this paper, we first introduce a quantum virtual machine named Qsun, whose operation is underlined by quantum state wavefunctions. The platform provides native tools supporting variational quantum algorithms. Especially using the parameter-shift rule, we implement quantum differentiable programming essential for gradient-based optimization. We then report two tests representative of quantum machine learning: quantum linear regression and quantum neural network.\end{abstract}
%
%
\maketitle


\section{Introduction}
The advent of quantum computers has opened 
a significant turning point for exponentially
speeding up computing tasks that classical computers need thousand years to execute~\cite{Arute2019, ChinaSupremacy}.
Although mankind has witnessed tremendous development in this field theoretically and experimentally in the last few years, most state-of-the-art quantum computers still rely on noisy intermediate-scale quantum computers NISQ. 
Noises and qubit-number constraints prevent 
to build high-fidelity quantum computers capable of 
substantially implementing quantum algorithms
\cite{Preskill2018quantumcomputingin,7927034,
Wang285,Kandala2019}. 
To bypass these constraints, various hybrid quantum-classical algorithms that use classical computers to optimize quantum circuits
have been proposed \cite{Cerezo2021, PhysRevA.101.010301, PhysRevA.103.L030401}. Among these, variational quantum algorithms (VQAs) may be the most promising ones in the NISQ era.

VQAs generally consist of three essential steps: (i) initializing quantum states for a given wavefunction ansatz, (ii) measuring a cost function suitable for problems being considered, and (iii) minimizing the cost function and updating new parameters. The self-consistency is performed until convergence.  
VQAs have been extensively employed to tackle numerous tasks, 
including the variational quantum eigensolvers (VQEs)
\cite{Peruzzo2014,
PhysRevResearch.1.033062,
Kirby2021contextualsubspace,
Ryabinkin2019,
PhysRevA.101.052340,
Gard2020,
PRXQuantum.2.020337}, 
quantum dynamics simulation
\cite{PhysRevX.7.021050,McArdle2019,
Nishi2021,PRXQuantum.2.030307,
zhang2020lowdepth,
PhysRevResearch.3.023095}, 
mathematical applications
\cite{PhysRevA.103.052425,
huang2019nearterm,
PhysRevLett.122.060504,
PhysRevA.103.052416,
bravoprieto2020variational,
xu2019variational,
PRXQuantum.2.010315,
Lloyd2014,
LaRose2019},
quantum machine learning (QML)
\cite{PhysRevA.98.032309,
farhi2018classification,
PhysRevLett.122.040504,
Havlicek2019,
PhysRevA.101.032308,
farhi2018classification,
Cong2019,
pesah2020absence,
zhang2020trainability,
Beer2020},
and new frontiers in quantum foundations
\cite{Arrasmith2019,wilde_2017,
Cerezo2021,
PhysRevA.101.062310,Koczor_2020,
PhysRevLett.123.260505,
beckey2020variational,
Meyer2021}. 

Typically, VQAs employ variational quantum circuits to measure the cost function on a quantum computer and outsource its optimization to a classical computer. While one can manipulate gradient-free optimizers, such as Nelder-Mead simplex \cite{Mead}, to minimize the cost function, using gradient-based methods like gradient descent can help us speed up and guarantee the convergence of the optimization. Several quantum algorithms have been proposed to evaluate the cost function gradient measured on quantum computers \cite{10.1145/3385412.3386011,Sch_fer_2020,Stokes2020quantumnatural,lopatnikova2021quantum,yamamoto2019natural,PhysRevLett.126.140502}. Among those methods, quantum differentiable programming (QDP) has been introduced and utilized extensively 
\cite{10.1145/3385412.3386011,Sch_fer_2020,guerreschi2017practical,
bergholm2020pennylane,
PhysRevLett.118.150503, 
PhysRevA.98.032309,PhysRevA.99.032331,
Banchi2021measuringanalytic}. 
It relies on a technique called 
the parameter-shift rule that evaluates 
the derivative of any differentiable function 
using quantum circuits 
\cite{guerreschi2017practical,
bergholm2020pennylane,
PhysRevLett.118.150503, 
PhysRevA.98.032309,PhysRevA.99.032331,
Banchi2021measuringanalytic}. Therefore, this method is beneficial for 
developing “on-circuit” gradient-based 
optimization techniques, especially for 
quantum machine learning (QML) applications 
where various methods like quantum neural networks 
(QNNs) demand the derivative information of the cost function.

While quantum algorithms 
should be performed 
on quantum computers, 
the current limitation of 
NISQ computers cause challenges 
in developing and testing 
new quantum algorithms, 
demanding the use of 
virtual alternatives 
called quantum virtual 
machines (QVMs).
Besides, 
QVMs are necessary for modeling 
various noisy channels to characterize 
the noises and the efficiency 
of quantum error correction.
One can classify QVMs into two types according to the way to build them: (i) the matrix multiplication approach \cite{qiskit,cirq_developers_2021_5182845,smith2017practical,bergholm2020pennylane,7927034,10.1145,Villalonga2019,roberts2019tensornetwork},
and (ii) the wavefunction approach \cite{suzuki2020qulacs,Guerreschi_2020,Steiger2018projectqopensource,kelly2018simulating, Jones2019,DERAEDT2007121,DERAEDT201947}.
While the former performs matrix multiplication 
for all qubits in quantum circuits, 
the latter represents quantum circuits 
by corresponding wavefunctions. 
On the one hand, the former can significantly 
reduce the memory capacity 
using tensor network contraction 
\cite{doi:10.1137/050644756,PhysRevLett.125.060503}, 
and the cost to pay is 
exponentially increasing 
the computational time.
On the other hand, the latter, in principle, can require less computational time in a limit of qubits number and a sufficiently large memory size that can store the total quantum wavefunction.
Therefore, both approaches exist 
side by side 
and a possible hybrid approach
\cite{Arute2019,markov2018quantum}
for convenient purposes.   

There are several QVM's libraries 
developed for QML orientation, such as
TensorFlow Quantum library  
\cite{broughton2020tensorflow} implemented in Cirq 
\cite{cirq_developers_2021_5182845},
Pennylane \cite{bergholm2020pennylane}
designed for photonics devices,
and TensorNetwork \cite{roberts2019tensornetwork,Huggins_2019}.
These libraries are all constructed in 
the matrix-multiplication type of QVMs, 
designed to submit quantum tasks 
to the developed hardware conveniently.

In this work, 
we develop a QVM platform 
named Qsun using the wavefunction approach
towards the QML applications. 
In Qsun, a quantum register is represented 
by a wavefunction, and quantum gates 
are manipulated directly by updating 
the amplitude of the wavefunction. 
Measurement results 
rely on probabilities of the wavefunction. Our simple approach yields faster computation speed for a small number of qubit when compared to other QVMs such as Qiskit, ProjectQ, or Pennylane. Basing on this generic QVM, 
we aim to exploit the advantages of QDP 
with the parameter-shift rule as the core engine towards practical applications 
in quantum machine learning. Two representative
examples of QML are demonstrated: quantum linear regression and quantum neural network. These algorithms are compared to standard progams: Qiskit, ProjectQ, and Pennylane, and classical algorithms when applicable. In these comparisons, Qsun performs slightly better for QDP, QLR, and QNN. All in all, Qsun is an efficient combination of QVM with QDP features that is oriented toward machine learning problems. In the following, we introduce the QVM platform Qsun and its performance compared to others in Sec.~\ref{secii}, followed by an introduction to the QDP implementation within Qsun in Sec.~\ref{seciii}. We then discuss some QML applications of the Qsun package in Sec.~\ref{seciv}.

\section{Quantum virtual machine implementation in Qsun}
\label{secii}
In practice, quantum computers work on quantum algorithms by  composing a quantum register, or qubits, operated by a sequence of quantum gates. Results are then traced out from quantum measurements. 
We now introduce our quantum virtual machine (QVM) named Qsun, an open-source platform simulating the operation of a generic quantum computer \cite{QSUN}.
We aim the platform to the development of quantum machine learning (QML) and related problems. We develop it in Python and employ the Numpy library 
for fast matrix operations and numerical computations. 

\subsection{Simulating quantum computers using wavefunction basis}
Unlike widely-used approaches based on matrix 
multiplication \cite{qiskit,cirq_developers_2021_5182845,smith2017practical,bergholm2020pennylane,7927034,10.1145,Villalonga2019,roberts2019tensornetwork}, our platform is developed using the class of ``wavefunction'' approach \cite{suzuki2020qulacs,Guerreschi_2020,Steiger2018projectqopensource,kelly2018simulating, Jones2019,DERAEDT2007121,DERAEDT201947}, in which a quantum register is represented by its wavefunction. The 
operation of quantum gates is simulated by updating the wavefunction's amplitude, and output results are obtained by measuring  wavefunction's probabilities. We expect that working directly on wavefunction is beneficial for QML applications, especially for building and training variational quantum circuits in quantum neural networks (QNNs).
As depicted in Fig.~\ref{fig:class},  Qsun consists of three main modules \texttt{Qwave, Qgates}, and \texttt{Qmeas} for quantum register, quantum gates, and quantum measurement, respectively.

\subsubsection*{Qwave}

\begin{figure}[t!]
\centering
\includegraphics[width=8cm]{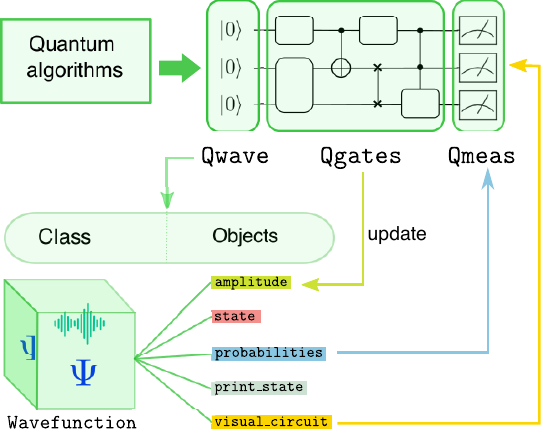}
\caption{Qsun quantum virtual machine (QVM) consists 
of three main modules \texttt{Qwave}, 
\texttt{Qgates}, and \texttt{Qmeas}, 
to simulate quantum circuits with 
a quantum register (qubits), 
quantum gates, and quantum measurements, 
respectively. 
In Qsun, a quantum register 
is constructed by its wavefunction through the 
\texttt{Wavefunction} class and 
its containing objects, as illustrated in the figure.}
\label{fig:class}
\end{figure}

In general, for a quantum register with $N$ qubits, its quantum states are represented in the $2^N$-dimension Hilbert space as 
\begin{align}\label{eq:Nqubit}
|\psi\rangle = \sum_{j = 0}^{2^N-1}\alpha_j |j\rangle,
\end{align}
where $\alpha_j$ are complex amplitudes obeying a completeness relation
$\sum_{j =0}^{2^N-1}|\alpha_j|^2 = 1$, and vectors $|j\rangle$ are 
elements of the computational basis. 
We integrate quantum state's information into the class \texttt{Wavefunction} as described in Fig.~\ref{fig:class} and Table.~\ref{t:1}.
The class allows us to access and update amplitudes directly according to 
the evolution of the quantum state under the action of the unitary quantum gates.  It also measures probabilities that contain output information.

\begin{table}
\small
\begin{center}
\caption {
List of methods used in the class \texttt{Wavefunction}.} \label{t:1}
    \begin{tabularx}{\linewidth}{@{}p{3cm}Xc>{\centering\arraybackslash}p{2cm}c@{}}
        \hline
        \textbf{Methods} & \textbf{Description} \\
        \hline
    \texttt{amplitude} & a NumPy array of complex numbers that stores the amplitudes of quantum states. \\
   \rowcolor[gray]{.9} \texttt{state} & a NumPy array of strings that labels the quantum states's basis. \\
    \texttt{probabilities} & return a list of corresponding probabilities for each basis vector in the superposition. \\
   \rowcolor[gray]{.9} \texttt{print\_state} & return a string representing a quantum state of the system in bra-ket notations. \\
  \texttt{visual\_circuit} & print a visualization of a quantum circuit. \\
    \hline
    \end{tabularx}
\end{center}
\end{table}

\subsubsection*{Qgates}

To manipulate a single-qubit gate $U$ 
$=\begin{pmatrix} a & c \\
b & d \end{pmatrix}$ 
acting on the $n^{\rm th}$ qubit, the nonzero elements in \texttt{amplitude} array are updated as \cite{Jones2019}
\begin{align}\label{eq:gate}
\begin{pmatrix}
\alpha_{s_i}\\ 
\alpha_{s_i+2^n}
\end{pmatrix}
\to U
\begin{pmatrix}
\alpha_{s_i}\\ 
\alpha_{s_i+2^n}
\end{pmatrix},
\end{align}
where $s_i = {\rm floor}(i/2^n)2^{n+1}+(i \text{ mod } 2^n)$, 
for all $i \in [0, 2^{N-1}-1]$. Here, ${\rm floor}(x)$ is the standard floor function taking the greatest integer less than or equal to the real of $x$. We unify the implementation of single- and multiple-qubit gates into a common framework. 
We outline the operation of single-qubit gates in Algorithm~\ref{al:1} and an example of the Hadamard gate in Algorithm~\ref{al:hadamard}. We emphasize that \texttt{Qgates} only update nonzero components of wavefunction amplitudes.
This way, we can avoid demanding matrix multiplications that escalate exponentially ($2^N\times 2^N$) with the number of $N$ qubits. This generic algorithm allows us to implement 
arbitrary unitary gates without decomposing them into universal ones,
which may be advantageous to model a general class of neural networks using quantum circuits.
\begin{algorithm}
\SetAlgoLined
\KwResult{Wavefunction with new probability amplitudes}
 w $\leftarrow$ Wavefunction\;
 states $\leftarrow$ w.state;
 ampl $\leftarrow$ w.amplitude\;
 $N$ $\leftarrow$ size(state[$i$][0]);
 $n$ $\leftarrow$ target qubit\;
 $n$\_ampl $\leftarrow$ [0, ..., 0], size($n$\_ampl) = size(ampl)\;
 $cut \leftarrow$ $2^{N-n-1}$\;
 \For{i $\leftarrow$ 0, {\rm size(ampl)}}{
  \eIf{{\rm state[$i$][$n$] == 0}}{
   $n$\_ampl[$i$] $\leftarrow$ $n$\_ampl[$i$] + $a$*ampl[$i$]\;
   $n$\_ampl[$i+cut$] $\leftarrow$ $n$\_ampl[$i+cut$] + $b$*ampl[$i$]\;
   }{
   $n$\_ampl[$i$] $\leftarrow$ $n$\_ampl[$i$] + $d$*ampl[$i$]\;
   $n$\_ampl[$i-cut$] $\leftarrow$ $n$\_ampl[$i-cut$] + $c$*ampl[$i$]\;
  }
 }
 w.amplitude $\leftarrow n$\_ampl
 \caption{
 Operation of a single-qubit gate:}
 \label{al:1}
\end{algorithm}

\begin{algorithm}
\SetAlgoLined
\KwResult{Wavefunction with new probability amplitudes corresponding to Hadamard state}
 w $\leftarrow$ Wavefunction\;
 states $\leftarrow$ w.state;
 ampl $\leftarrow$ w.amplitude\;
 $N$ $\leftarrow$ size(state[$i$][0]);
 $n$ $\leftarrow$ target qubit\;
 $n$\_ampl $\leftarrow$ [0, ..., 0], size($n$\_ampl) = size(ampl)\;
 $cut \leftarrow$ $2^{N-n-1}$\;
 \For{i $\leftarrow$ 0, {\rm size(ampl)}}{
  \eIf{{\rm state[$i$][$n$] == 0}}{
   $n$\_ampl[$i$] $\leftarrow$ $n$\_ampl[$i$] + $(1/\sqrt{2})$*ampl[$i$]\;
   $n$\_ampl[$i+cut$] $\leftarrow$ $n$\_ampl[$i+cut$] + $(1/\sqrt{2})$*ampl[$i$]\;
   }{
   $n$\_ampl[$i$] $\leftarrow$ $n$\_ampl[$i$] + $(-1/\sqrt{2})$*ampl[$i$]\;
   $n$\_ampl[$i-cut$] $\leftarrow$ $n$\_ampl[$i-cut$] + $(1/\sqrt{2})$*ampl[$i$]\;
  }
 
 w.amplitude $\leftarrow n$\_ampl
 \caption{
 Operation of Hadamard gate:}}
 \label{al:hadamard}
\end{algorithm}

To mimic the actual operation of quantum computers, 
we introduce noises into the wavefunctions. 
In Qsun, the standard quantum depolarizing channel is implemented 
as a single-qubit gate $\mathcal{E}$ that is a part of 
quantum circuits acting on the wavefunctions.
For a given noisy probability $p$,  
applying the gate $\mathcal{E}$
on a mixed quantum state $\rho$ 
will transform it to~\cite{nielsen_chuang_2019} 
\begin{align}\label{eq:eg}
\rho \xrightarrow{\mathcal{E}}
(1-p)\rho+\dfrac{p}{2}I,
\end{align}
where $I$ is an $2\times2$ identity matrix.
In general, one can decompose the depolarizing channel
into the bit-flip, phase-flip, and phase-bit-flip as
\begin{align}\label{eq:noise}
\rho \xrightarrow{\mathcal{E}}
(1-p)\rho 
+ p_x X\rho X
+ p_y Y\rho Y
+ p_z Z\rho Z,
\end{align}
where $p_x, p_y, p_z$ are the 
probabilities of bit-flip, phase-bit-flip,
and phase-flip, respectively 
\cite{nielsen_chuang_2019,7927034}.
In Qsun, we use $p_x = p_y = p_z = p/3$
and apply to every qubits in 
the circuit each after the action of a quantum gate
on the circuit.

\subsubsection*{Qmeas}

The module \texttt{Qmeas} is designed 
to execute quantum measurements on a single qubit 
or all qubits in the quantum circuit.
For a measurement on a single qubit $n$,
the probability for that its outcome is $|0\rangle$ reads
\begin{align}\label{eq:prob}
 p(0) = \sum_{i=0}^{2^{N-1}-1}\langle\psi|j_{s_i}\rangle\langle j_{s_i}|\psi\rangle
=\sum_{i=0}^{2^{N-1}-1}|\alpha_{s_i}|^2,
\end{align}
with $|j_{s_i}\rangle$ as the basis element and the post-quantum state after the measurement
is given as
\begin{align}\label{eq:poststate}
|\psi'\rangle =\dfrac{\sum_{i=0}^{2^{N-1}-1}
|j_{s_i}\rangle\langle j_{s_i}|\psi\rangle}
{\sqrt{p(0)}}.
\end{align}
Similarly, the probability 
for getting the outcome $|1\rangle$
reads
\begin{align}\label{eq:prob1}
 p(1) = 1-p(0), 
\end{align}
and the post-state is
\begin{align}\label{eq:poststate1}
|\psi'\rangle =\dfrac{\sum_{i=0}^{2^{N-1}-1}
|j_{s_i+2^n}\rangle\langle j_{s_i+2^n}|\psi\rangle}
{\sqrt{p(1)}}.
\end{align}

For all-qubit measurement, 
the post-quantum state will
collapse to one of $\{|j\rangle\}$
with the probability of $|\alpha_j|^2$.
In Qsun,
we build these two measurements
onto \texttt{measure\_one} and 
\texttt{measure\_all}, respectively. 

\begin{figure}[t]
\centering
\includegraphics[width=8.0cm]{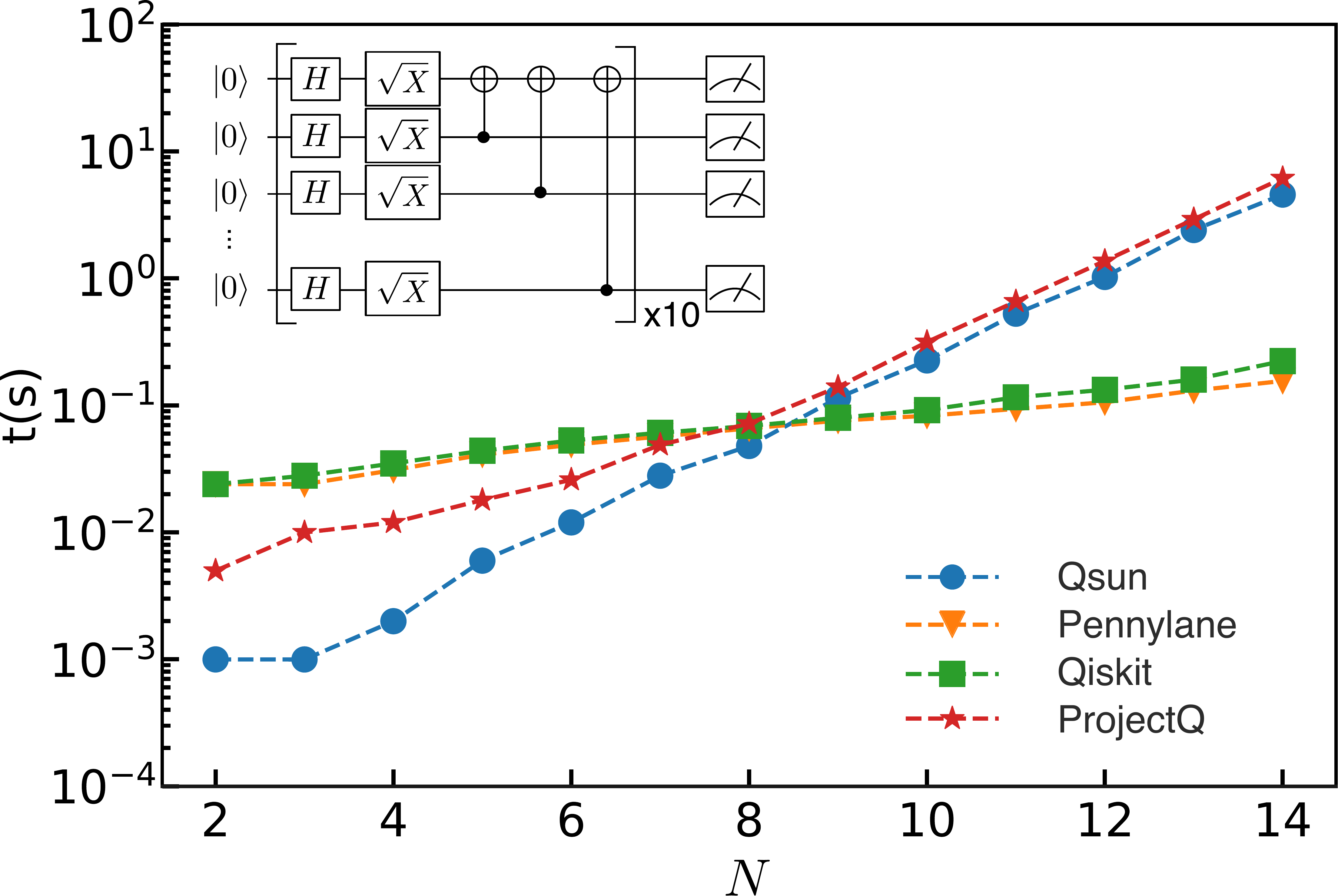}
\caption{Comparing computational time $t$ (in seconds) of different QVMs as increasing the number of qubits $N$. Inset: The testing circuit including an $H$, a $\sqrt X$, and a CNOT gates acts on each qubit. The depth is fixed at 10, and the number $N$ of qubits is varied to assess the QVM performance.}
\label{fig:benchmarking1}
\end{figure}

\subsection{Assessing the QVM performance}

Let us now assess the performance of Qsun and compare it with three existing ones in Pennylane, Qiskit, and ProjectQ.
Note that Qsun and ProjectQ  belong to the wavefunction class, while the others are in the matrix multiplication class.
We have adopted the testing circuit from Ref.~\cite{quantest} composed of the Hadamard, $\sqrt{X}$, and CNOT gates acting on each qubit.
We have fixed the depth of the circuit at 10 and varied the number of qubits $N$. The code for this test is shown in \ref{appendix:test}.

Fig. \ref{fig:benchmarking1} represents the change of computational time when the number $N$ of qubits increases. In general, there are two magnitudes of slope corresponding to two ways of QVM implementation. While the wavefunction-based approach is faster than the matrix-multiplication one for small numbers of qubits ($N < 8$), the opposite behavior is observed for larger numbers of qubits.
This observation reflects the basic properties of these two approaches as discussed above and see also Ref. \cite{suzuki2020qulacs}.
However, there are some available techniques to improve the performance of the wavefunction approach for larger numbers of qubits, such as, SIMD (single-instruction, multiple data) optimization and multi-threading \cite{suzuki2020qulacs}. 
We further summarize a comparison between Qsun and other simulators in terms of 
practical quantum algorithms in Table~\ref{t:2}.

\begin{widetext}
\begin{center}
\begin{table}
\small
\caption {Comparison between Qsun and other simulators} \label{t:2}
    \begin{tabular}{|c|c|c|c|c|}
        \hline
        \textbf{Algorithms} & \textbf{Qsun} & \textbf{ProjectQ} & \textbf{Qiskit} & \textbf{Pennylane} \\
        \hline
    Standard algorithms & \textcolor{green}\cmark & \cmark & \cmark & \cmark \\
   \rowcolor[gray]{.9} Quantum diferentiable programming & \textcolor{green}\cmark & \cmark & \cmark & \cmark \\
   Quantum Linear Regression  & \textcolor{green}\cmark & \cmark & \cmark & \cmark  \\
   \rowcolor[gray]{.9} Quantum Neural Network  & \textcolor{green}\cmark & & \cmark & \cmark \\
    \hline
    \end{tabular}
\end{table}
\end{center}
\end{widetext}

\section{Quantum differentiable programming implementation in Qsun}
\label{seciii}

Given a quantum state $|\psi(\vec\theta)\rangle$ with $\vec\theta$ as variational parameters and an observable $\widehat C$,
the task is to seek the global minimum of 
the expectation value 
$C(\vec\theta) = \langle\psi(\vec\theta)|\widehat C|\psi(\vec\theta)\rangle$ with respect to parameters $\vec\theta$. For example, if $\widehat C$ is a Hamiltonian, its global minimum is the ground state energy. In general, $C(\vec\theta)$ is called as the cost function, and minimizing the cost function requires its derivative with respect to parameters $\vec\theta$, $\partial C(\vec\theta) / \partial\theta$. In classical computing, if the analytical form of $\partial C(\vec\theta) / \partial\theta$ is unknown, finite difference methods are often used to evaluate the derivative approximately. Although this approximation is fast and easy to implement, its accuracy depends on discretization steps. In contrast to the classical finite differentiation, quantum differentiable programming (QDP) is an automatic and exact method to compute the derivative of a function. 
QDP is thus essential for accurate gradient computation in multiple VQAs, including QML models.

\begin{algorithm}
\SetAlgoLined
\KwResult{Derivative of a function}
 $f \leftarrow$ Function;
 $c \leftarrow$ Quantum Circuit\;
 $p \leftarrow$ Params;
 $s \leftarrow$ Shift\;
 {\it diff} $\leftarrow$ [0, ..., 0], size({\it diff}) = size($p$)\;
 \For{i $\leftarrow$ 0, {\rm size($diff$)}}{
    $p$\_plus $\leftarrow$ copy($p$)\;
    $p$\_subs $\leftarrow$ copy($p$)\;
    $p$\_plus[$i$] $\leftarrow$ $p[i] + s$\;
    $p$\_subs[$i$] $\leftarrow$ $p[i] - s$\;
    {\it diff}[$i$] $\leftarrow (f(c, p$\_plus) $- f(c, p$\_subs))/(2*sin($s$))\;
 }
 \caption{
 Quantum differentiable programming implementation in Qsun}
\label{al:QDP}
\end{algorithm}

The heart of QDP is the parameter-shift rule that is analytically computed using
quantum circuits. The algorithm is outlined in Algorithm \ref{al:QDP}. Let us introduce a parameterized generator $\widehat V$ independent of $\vec\theta$ such that $|\psi(\vec\theta)\rangle = 
e^{i\vec\theta\widehat V}|\psi\rangle$.
The cost function is then rewritten as
\begin{align}\label{eq:cost}
  C(\vec\theta) = \langle\psi|e^{-i\vec\theta\widehat V}
    \widehat C e^{i\vec\theta\widehat V}|\psi\rangle
    ={\rm Tr}\bigl(\widehat C e^\mathcal{Z}[\rho]\bigr),
\end{align}
with $\rho = |\psi\rangle\langle\psi|$;
$\mathcal{Z} = i\vec\theta \widehat V$, 
and the superoperator $e^\mathcal{Z}[\rho]=
e^\mathcal{-Z}\rho e^\mathcal{Z}$
\cite{miller_1990}.
The parameter-shift rule for each $\theta\in\vec\theta $ states that
\begin{align}\label{eq:psr}
    \dfrac{\partial C(\vec\theta)}{\partial \theta}  
    ={\rm Tr}\bigl(\widehat C
    \partial_\theta e^\mathcal{Z}[\rho]\bigr)
    = c\bigl[ C(\vec\theta + s) - C(\vec\theta - s) \bigr],
\end{align}
where $c=1/(2\sin(s))$, and $s$ is determined based on the superoperator and independent of $\vec\theta$. 
The values of the cost function $C$ 
at $\vec\theta \pm s$ are measured on 
quantum computers by implementing 
two quantum circuits as follows
\begin{figure}[hbt!]
    \centering
    \leavevmode
    \Qcircuit @C=0.5em @R=1em {
    \lstick{\ket{\psi}} & \gate{e^{i(\vec\theta + s)\widehat V}} & \meter{C} & \rstick{C(\vec\theta + s)} \qw \\
    & &  \langle\widehat C\rangle  \\
    \lstick{\ket{\psi}} & \gate{e^{i(\vec\theta - s)\widehat V}} &\meter{C} & \rstick{C(\vec\theta - s)} \qw \\
    }
\end{figure}\\

The derivative is finally obtained by subtracting measurement results from the two circuits. An advantage of the parameter-shift rule is that it can compute the derivative of the given function exactly while the shift $s$ can be chosen arbitrarily large \cite{quanshiftrule}. 

\begin{figure}
\centering
\includegraphics[width= 8.0cm]{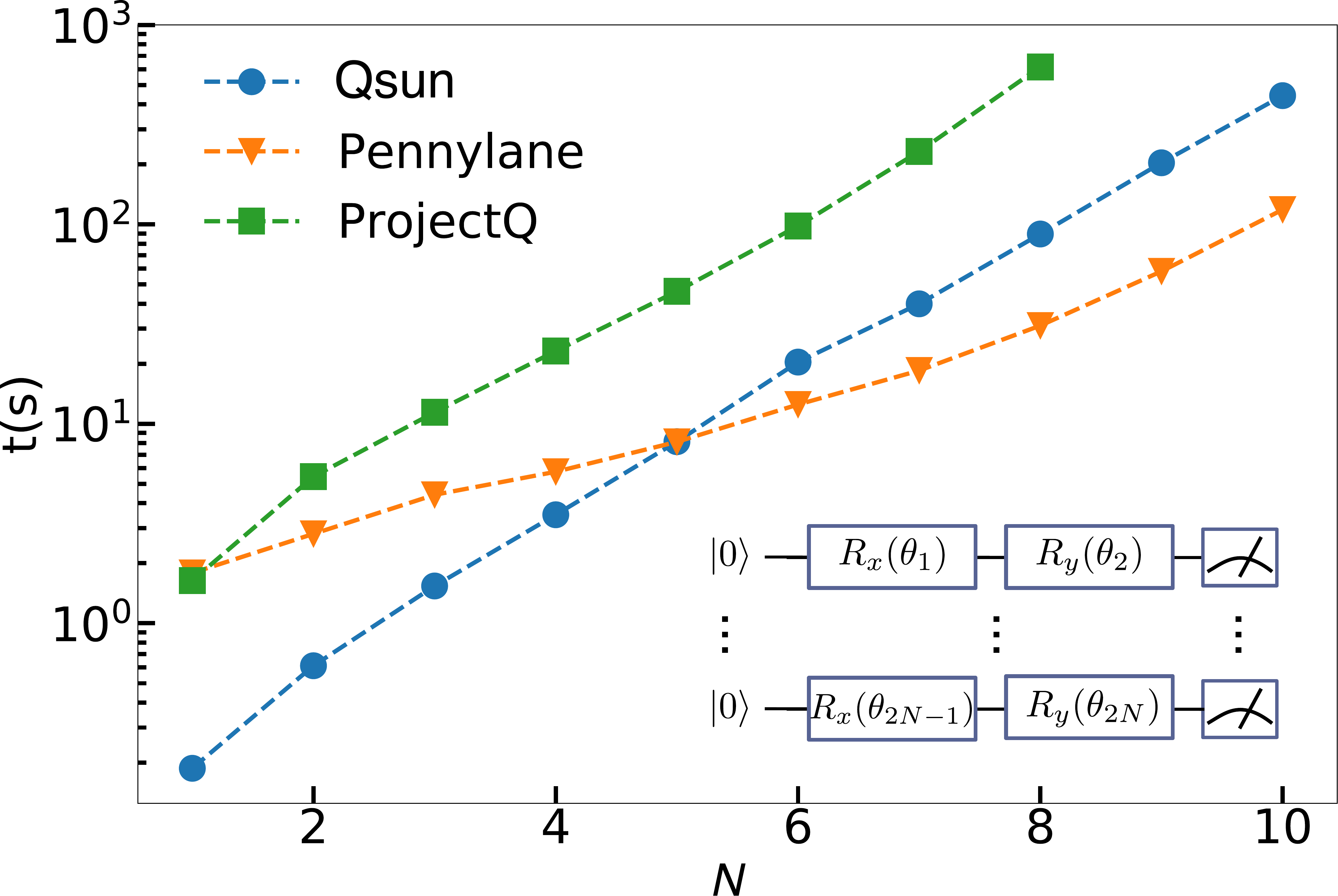}
\caption{Computational time $t$ in seconds of the QDP implementation in Qsun, ProjectQ using an engine Simulator() and Pennylane using a default.qubit device when the number of qubits increases. Inset: quantum circuits used for the QDP test.} \label{fig:test_qdp}
\end{figure}

We combine the QDP implemented in Qsun 
with a classical optimizer like adaptive moment (Adam) optimization 
and implement quantum gradient descent, 
making the optimization procedure fully quantum. 
Let us now examine the performance of the QDP 
in comparison with Pennylane, 
another QML-oriented platform. 
In this test, we measure 
the computational time spent for 
optimizing the ground state 
of an Ising model. 
The code for running this test 
is given in \ref{app_QDP}, 
and a tutorial explaining 
how to use QDP with sample codes is written
in \ref{appendix:tutorial}.

Fig~\ref{fig:test_qdp} presents 
the change of computational time 
when increasing the number of qubits $N$
for three cases of Qsun, Pennylane, and ProjectQ. 
As shown in the inset of Fig.~\ref{fig:test_qdp}, 
for each qubit, we have two variational parameters 
$\vec\theta$, meaning that the number of variational 
 parameters is twice of $N$. 
Qsun is faster than Pennylane for 
$N < 5$, whereas it is slower for larger numbers 
of qubits ($N > 5$). 
Among the three cases,
ProjectQ consumes the most computational time for all $N$.
Of course, the exponential growth with 
the number of qubits is not inevitable. 

\section{Quantum machine learning applications using  Qsun}
\label{seciv}
Various quantum machine learning models 
can be developed with quantum differentiable programming (QDP) implementations 
to evaluate the gradient and employ 
gradient-based optimization. 
This section demonstrates QML applications 
using Qsun in two well-known models: 
quantum linear regression (QLR) 
and quantum neural network (QNN). 
Its performances are compared 
to other standard tools.

Before digging into detailed examples,
let us derive the QDP implementation
in derivative of a standard mean squared 
error cost function:
\begin{align}\label{eq:mse}
    C(\vec \theta) = \frac{1}{M}
    \sum_{i=0}^{M-1} \Big(y_i - \widehat{y}_i\Big)^2,
\end{align}
where $\vec\theta$ is a tuple of 
variational parameters, $y_i$ represents 
the true value and $\widehat y_i$ 
stands for prediction value,
with $i\in [0, M-1], M$ 
the number of samples in the dataset to be trained.
Concretely, we also consider the prediction value
is a composition function of an activation function 
$f(\vec\theta)$, i.e., $\widehat y_i(f(\vec\theta))$.
Then, we derive at a chain rule:
\begin{align}\label{eq:gradchain}
    \frac{\partial C(\vec{\theta})}{\partial\theta}
    =\frac{1}{M}\sum_{i=0}^{M-1}
    \frac{\partial C(\widehat y_i)}{\partial\widehat y_i}
    \frac{\partial \widehat y_i(f)}{\partial f}
    \frac{\partial f(\vec{\theta})}
    {\partial \theta},
\end{align}
where $\theta\in\vec\theta$, with
\begin{align}
\frac{\partial C(\widehat y_i)}{\partial\widehat y_i}
    =-2[y_i-\widehat y_i],
\end{align}
the derivative $\frac{\partial \widehat y_i(f)}{\partial f}$
depends on the particular form of the activation function,
and 
\begin{align}\label{eq:gradxf}
\notag \frac{\partial f(\vec{\theta})}{\partial\theta}
    &=\sum_k \frac{\partial m_k(\vec\theta)}
    {\partial\theta}\\
    &=
    \sum_k c_k\Bigl[m_k(\vec\theta+s_k) 
    - m_k(\vec\theta-s_k)\Bigr],
\end{align}
where $m_k(\vec\theta) = \langle\psi(\vec\theta)| 
\mathcal{\widehat M}_k |\psi(\vec\theta) \rangle$
is a measurement outcome of an operator
$\mathcal{\widehat M}_k$ for a measurement set $\{k\}$.
To arrive at the second equality in Eq.~\eqref{eq:gradxf}, 
we have applied the parameter-shift rule \eqref{eq:psr}. 

\subsection{Quantum linear regression}

\begin{figure}[t]
    \centering
    \includegraphics[width=8.0cm]{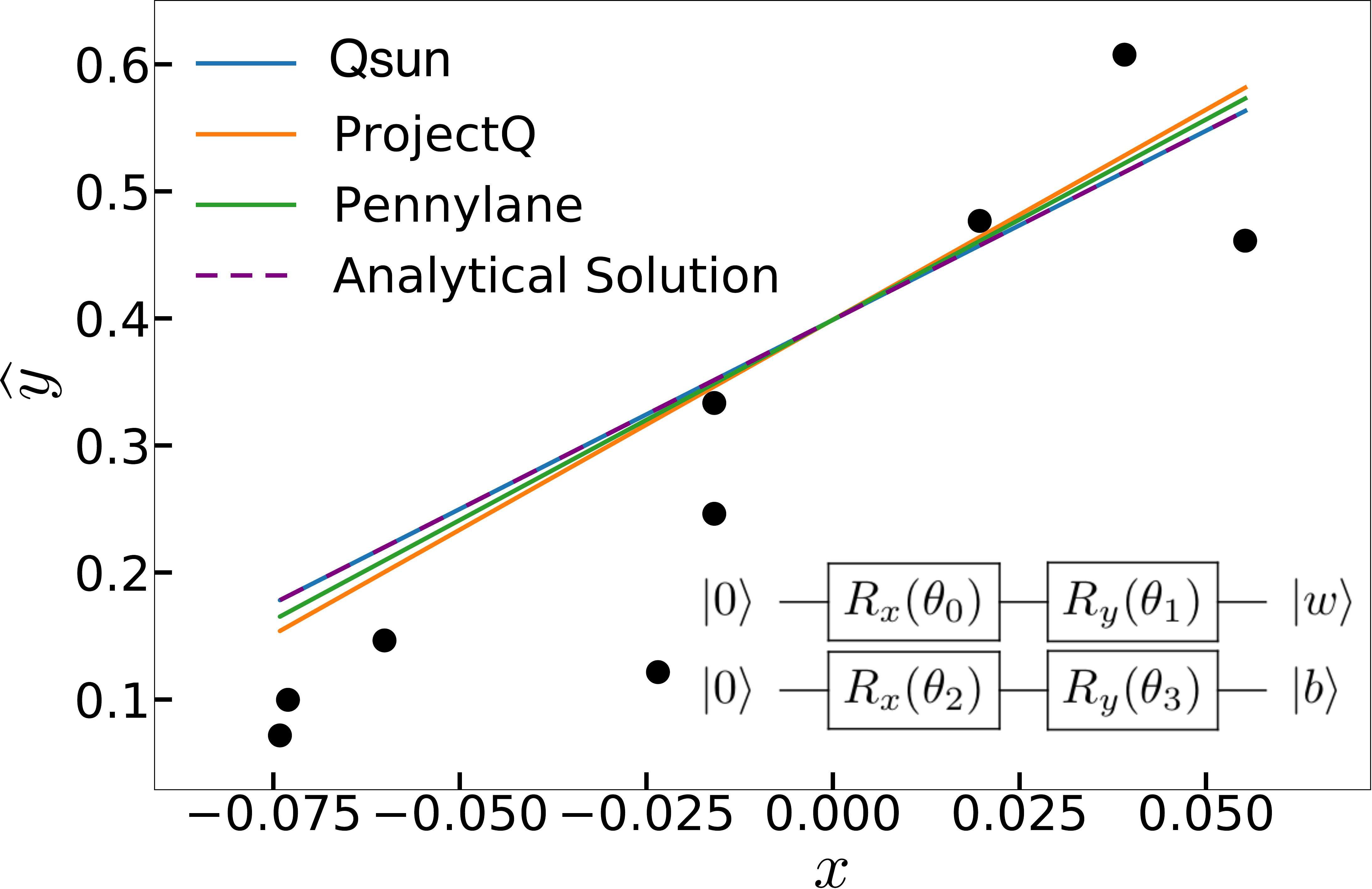}
    \caption{Linear regression models trained by quantum programming using Qsun (blue), Pennylane (green), ProjectQ (orange) 
    and analytical solution (dashed purple). 
    We use the diabetes dataset 
    \cite{diabetes} with 400 samples 
    for the training set and 
    10 samples for the testing set. 
    The boundary $k$ = 10 with 1000 iterations. 
    Inset: the quantum circuit that 
    trains the quantum linear regression models.} 
    \label{fig:compare_linear_regression}
\end{figure}

We implement quantum linear regression 
on a ``diabetes'' dataset \cite{diabetes} 
available in the scikit-learn package \cite{scikit-learn} 
with 400 samples for the training set 
and 10 samples for the testing set.
We write the linear regression model in the form
\begin{equation}
    \widehat{y} = w x + b ,
    \label{classic_linear_model}
\end{equation}
where $w$ and $b$ are the slope 
and intercept of the linear regression
need to be obtained. 
To evaluate them on quantum computers, 
we store their values in two qubits. 
The values of  $w$ and $b$ now become 
the expectation values of 
the Pauli matrix $\widehat\sigma_z$. 
The quantum version of 
linear regression model 
\eqref{classic_linear_model} states
\begin{equation}\label{linear_model}
    \widehat{y}(\vec\theta) = 
    k\big(\langle \widehat w \rangle x 
    + \langle \widehat b \rangle\big), 
\end{equation}
where $k$ is the scaling factor 
that transforms the output data 
from $[-1, 1]$ into $[-k, k]$.
Its value should be chosen so that the R.H.S can cover 
all of the true values $\{y_i\}$.
Here, $w, b$ are functions of 
variational parameters $\vec\theta$,
for example $w \equiv w(\vec\theta)  
= \langle \psi(\vec\theta)|\widehat\sigma_z |\psi(\vec\theta) \rangle$ 
with the initial state $|\psi\rangle = |0\rangle$,
(see the inset Fig.~\ref{fig:compare_linear_regression}).
The cost function is the mean squared error 
as described in Eq.~\eqref{eq:mse}.

After having the model, we run the circuit
in the inset Fig.~\ref{fig:compare_linear_regression}
to find the optimized values for $w$ and $b$
via updating $\vec\theta$.
We first calculate their derivatives using 
the chain rule as shown in 
Eq.~\eqref{eq:gradchain} and 
the parameter-shift rule in the QDP as
depicted in Algorithm \ref{al:QDP},
and then update new parameters
using gradient descent
\begin{align}\label{eq:grad-des}
    \theta_{\rm t+1} = \theta_{\rm t} -\eta
    \frac{\partial C(\vec\theta_{\rm t})}{\partial\theta},
\end{align}   
where $\eta$ the learning rate, 
$\vec{\theta}_{\rm t}$ is the set of parameters at time step t.
They are continuously updated 
until the optimized values are found. 
The quantum circuit that used to train the model 
is represented in the inset 
Fig. \ref{fig:compare_linear_regression}, 
and the code is given in \ref{appendix:regression}.

For comparison, we next provide 
the analytical solution for minimizing 
the cost function \eqref{eq:mse} 
with \eqref{classic_linear_model}. 
We vectorize:
\begin{align}
    \begin{pmatrix}
     w \\
     b
    \end{pmatrix} = (X^{T}X)^{-1}X^{T}Y
\end{align}
where
\begin{align}
    X = \begin{pmatrix}
     x_{0} & 1 \\
     \vdots & \vdots \\
     x_{M-1} & 1
    \end{pmatrix}, \quad Y = \begin{pmatrix}
     y_{0} \\
     \vdots \\
     y_{M-1}
    \end{pmatrix},
\end{align}
and $T$ the transpose operator. 

The performance of Qsun is compared to those of 
Pennylane, ProjectQ, and analytical solution using 
the same set of parameters. 
We use the maximum number of 
iterations 1000 and $k = 10$. 
For Pennylane, we use 
GradientDescentOptimizer() 
with its default configuration. For ProjectQ, we use the same optimize algorithm as we have used for Qsun, which is shown in \ref{appendix:regression}.
As we can see from Fig. 
\ref{fig:compare_linear_regression}, 
Qsun's result is closer to 
the analytical result than the Pennylane and ProjectQ one, which implies 
a high performance of Qsun. 
In the same case of the ``wavefunction'' approach, 
Qsun has a better efficiency than ProjectQ in the speedup as shown in 
Fig. \ref{fig:benchmarking1} and Fig. \ref{fig:test_qdp}, 
and in the optimization process.
    
\subsection{Quantum neural network}

\begin{figure*}[hbt!]
    \centering
    \includegraphics[width=14cm]{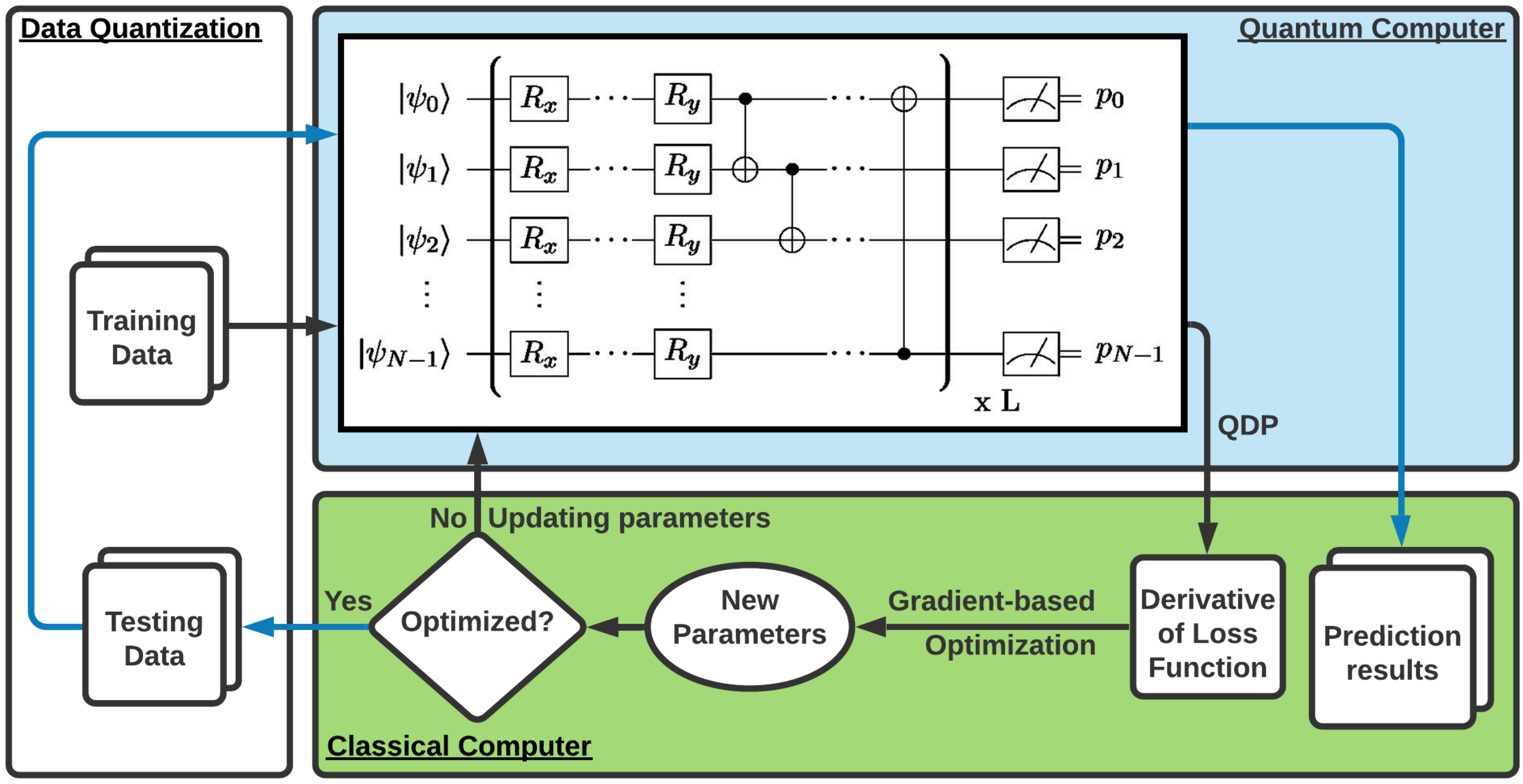}
    \caption{\label{fig:qnn_diagram.jpg}A Quantum Neural Network framework. The blue box is the quantum part of the QNN, and the green box is the classical part of the QNN. At first, quantify and encode the dataset (training and testing) into quantum states of qubits $|\psi_i\rangle$. Each feature in the dataset encodes into one qubit.
    The employed rotation gates will parameterize the quantum circuit, and the CNOT gates cause entangled in the circuit. These gates repeat L times for L layers in the quantum neural network. After that, we measure the circuit and give the corresponding probabilities $p_k$. We employ a  QDP scheme with the pentameter-shift rule to calculate the derivative of $p_k$ and send the results to a classical computer to derive the derivative of the loss function. After that, we implement a gradient-based optimization to obtain new parameters. When the scheme is not optimal yet, we update the circuit with new parameters; when it is optimal, 
    we turn it to the testing process.}
\end{figure*}

In this subsection, we show the ability of Qsun for deep learning by building and training a quantum neural network (QNN). We model the QNN as a variational quantum circuit (VQC) parameterized with multiple variables that are optimized to train the model. Fig. \ref{fig:qnn_diagram.jpg} represents 
the process for building and training our QNNs. It is a hybrid quantum-classical scheme with five steps summarized as follows: 

\begin{itemize}
        \item The quantum part:
        \item[]{Step 1:} Quantizing and encoding dataset into quantum states using the amplitude encoding method \cite{PhysRevA.101.032308,PhysRevA.102.032420}.
        \item[]{Step 2:} Building a QNN circuit and measurement.
        \item[]{Step 3:} Evaluating the derivative of measurement results using QDP.
        \item The classical part:
        \item[]{Step 4:}
        Deriving the derivative of the defined cost (loss) function.
        \item[]{Step 5:} Running a gradient-based optimization and updating parameters.
\end{itemize}
    
\subsubsection*{Data quantization and amplitude encoding}

Set $x_j^{(i)}$ as the $i^{\rm th}$ sample ($i\in[0, M-1]$) of the $j^{\rm th}$ feature ($j \in [0, N-1]$) in the dataset of $N$ features with $M$ samples for each feature. 
Since we only analyze one sample (other samples are treated similarly), we can omit the indicator $i$ and 
rewrite $x_j^{(i)}$ as $x_j$. We now map $x_j$ into a qubit by normalizing its value in range $[x_j^{\rm (min)}, x_j^{\rm (max)}]$ into $[0, 1]$. 
Using the Min-Max normalization, we 
obtain
\begin{align}
\widetilde x_j = \dfrac{x_j - x_j^{\rm (min)}}{x_j^{\rm (max)} - x_j^{\rm (min)}}.
\end{align}
Here, the normalized value $\widetilde x_j \in [0, 1]$.
We map this value into 
the amplitudes of a qubit as
$|\psi_j\rangle = \sqrt{\widetilde x_j} |0\rangle + \sqrt{1-\widetilde x_j} |1\rangle.$
Then, the quantum state for $N$ features reads
\begin{align}
    |\Psi\rangle = 
    \bigotimes^{N-1}_{j=0} \ket{\psi_{j}}
    =\bigotimes^{N-1}_{j=0}
    \left( \sqrt{\widetilde x_j}\ket{0} + \sqrt{1-\widetilde x_j} \ket{1} \right).
\end{align}
The encoding implementation is given in Algorithms \ref{al:Encoding}:

\begin{algorithm}
\SetAlgoLined
\KwResult{Probablity amplitudes of QNN's initial quantum states}
 sample $\leftarrow$ sample\_scaled\;
 $N$ $\leftarrow$ number\_of\_features\;
 ampl $\leftarrow$ [0, ..., 0], size(ampl) = $N$\;
 \For{i $\leftarrow$ 0, N-1}{
    ampl[$i$] $\leftarrow$ [sqrt(sample[$i$]), sqrt(1-sample[$i$])]\;
 }
 ampl\_vec $\leftarrow$ $\text{ampl[0]} \otimes \text{ampl[1]} \otimes \cdots \otimes \text{ampl[$N$-1]}$\;
 \caption{
 Encoding data in Qsun}
\label{al:Encoding}
\end{algorithm}

\subsubsection*{Building the QNN circuits}
The QNN model uses $N$ qubits for $N$ features 
and includes multiple layers. It is parameterized through a set of variables $\vec{\theta} = \{\theta_{kjj'}\}$ with $k$ as the layer index, $j$ as qubit (feature) indices,
and $j'$ indicates the number of rotation gates implementing on each qubit.
Each layer has one $R_x$ and $R_y$ 
gates for each qubit followed by CNOT gates 
generating all possible entanglements 
between them if the number of qubits 
is more than one. 

\subsubsection*{Decoding probabilities into predictions}

Here, we map measurement results from the previous step into classical predictions by using an activation function.
We consider the expectation value of the projection operator $\widehat\Pi = |1\rangle\langle 1|$ as
\begin{align}
    p^{(i)}_j\equiv
    \braket{\widehat\Pi} 
    = \langle\psi^{(i)}_j|1\rangle\langle
    1|\psi^{(i)}_j\rangle 
    = |\langle
    1|\psi^{(i)}_j\rangle|^2.
\end{align}
Note that $|\psi^{(i)}_j\rangle$ 
is the final state of qubit $j$
at sample $i$.
We also emphasize that 
one can choose 
the projection operator 
$\widehat \Pi$ arbitrarily. 
We use the sigmoid function, 
$S(x) = 1/(1+e^{-x})$ to transform 
the measurement data into predictions.
In our concrete example below, 
we have two features represented 
by two qubits, i.e, $j = 0, 1$.
We use the prediction rule as
\begin{align}
     S_i(p) = 
    S\bigl(\gamma(p^{(i)}_0-p^{(i)}_1)\bigr),
\end{align}
where $\gamma$ is a scaling shape for the 
sigmoid function such that for large $\gamma$
then $S(x)$ becomes a Heaviside step function.
To use a soft prediction, 
we introduce a label-conditional 
probability for a prediction value as
\begin{equation}
\widehat{y}_{i}(l) = \left\{\begin{array}{lr}
S_i(p), & \text{for } l = 0\\
1- S_i(p), & \text{for } l  = 1
\end{array}\right..
\end{equation}

\subsubsection*{Training the QNN model}
In the current version of Qsun, 
we are using a quantum-classical hybrid scheme 
combining QDP and the classical Adam optimization 
to train our QNN model. 
The cost function is defined by
\begin{equation}
        C(\vec{\theta}) = \frac{1}{M} 
        \sum_{i=0}^{M-1} 
        \bigl[ 1-\widehat{y}_{i}(y_i) \bigr]^{2}, \label{eq:cost2}
\end{equation} 
which is a function 
of $\vec{\theta}$. Here, $y_{i} \in \{0, 1\}$ are true values, 
$\widehat{y}_{i}(y_i)$ are predicted probability
conditioned on the label $y_i$. 
We then evaluate the derivative of the cost function using QDP, 
followed by a gradient-descent procedure to search for optimal parameters $\vec{\theta}$. 

The derivative of the cost function 
concerning each $\theta_{kjj'} \in \vec{\theta}$ 
following the chain rule \eqref{eq:gradchain}
gives
(hereafter, we omit its indices)
\begin{align}\label{eq:grad}
    \frac{\partial C(\vec{\theta})}{\partial\theta}
    =\frac{1}{M}\sum_{i=0}^{M-1}\frac{\partial C(y_i)}{\partial\widehat y_i(l)}
    \frac{\partial \widehat y_i(l)}{\partial p}
    \frac{\partial p(\vec{\theta})}
    {\partial \theta},
\end{align}
with
\begin{align}
\frac{\partial C(y_i)}{\partial\widehat y_i(l)}
    &=-2[1-\widehat y_i(l)], \\
\frac{\partial \widehat y_i(l)}{\partial p}
    &=(-1)^l\widehat y_i(l) [1-\widehat y_i(l)],
\end{align}
and
\begin{align}
\notag \hspace{-0.25cm}\frac{\partial p(\vec{\theta})}{\partial\theta}
    &=\gamma\frac{\partial}{\partial\theta}
    \Bigl(p^{(i)}_0-p^{(i)}_1\Bigr)\\
\notag   &=\gamma\Bigl\{
    c_0\bigl[p^{(i)}_0(\theta+s_0) - p^{(i)}_0(\theta-s_0)\bigr]\\
    &\hspace{0.5cm} 
    -c_1\bigl[p^{(i)}_1(\theta+s_1) - p^{(i)}_1(\theta-s_1)\bigr]
    \Bigr\},\label{eq:gradx}
\end{align}
here, we have applied the parameter-shift rule \eqref{eq:psr}. 
We finally update the model with new parameters using 
the Adaptive Moment (Adam) optimization algorithm~\cite{kingma2017adam} 
as follows. Let $g_{\rm t+1}$ 
be the gradient of the cost function w.r.t.  
$\theta$ at time step t+1, (t starts from 0):
\begin{align}\label{eq:gt1}
    g_{\rm t+1} \equiv \frac{\partial C(\vec{\theta}_{\rm t})}{\partial \theta},
\end{align}
where $\vec{\theta}_{\rm t}$ is the set of parameters at time step t.
Next, we estimate the first and second order moments of the gradient as 
\begin{align}\label{eq:fs}
    v_{\rm t+1} = \beta_{1} v_{t} + (1-\beta_{1}) g_{t+1} \\
    w_{\rm t+1} = \beta_{2} w_{t} + (1-\beta_{2}) g^{2}_{t+1},
\end{align}
and compute 
\begin{align}\label{eq:vwh}
    \widehat{v}_{\rm t+1} = \frac{v_{\rm t+1}}{1 - \beta_{1}^{t+1}}, 
    \text {and }
    \widehat{w}_{\rm t+1} = \frac{w_{\rm t+1}}{1 - \beta_{2}^{t+1}}.
\end{align}
Here, $\beta_1, \beta_2$ and $\epsilon$ are nonnegative weights, originally suggested as $\beta_1 = 0.9, \beta_2 = 
0.999$ and $\epsilon = 10^{-8}$.
Also, $v_0 = w_0 = 0$.
Finally, the parameter $\theta$ 
is updated to
\begin{align}\label{eq:update}
    \theta_{\rm t+1} = \theta_{\rm t} - \frac{\eta \widehat{v}_{\rm t+1}}{\sqrt{\widehat{w}_{\rm t+1}} + \epsilon},
\end{align}
where $\eta$ is the learning rate. 
The implementation of Adam algorthims in a combination with QDP is given in  Algorithm~\ref{al:GD}. 
\begin{algorithm}
\SetAlgoLined
\KwResult{Updated parameters}
 $c \leftarrow$ Quantum Circuit\;
 $p \leftarrow$ Params;
 $s \leftarrow$ Shift\;
 $\beta_{1}, \beta_{2}, \epsilon, \eta \leftarrow$ 
 Beta$_{1}$, Beta$_{2}$, Epsilon, Eta\;
 $v$\_adam, $s$\_adam $\leftarrow$ $v$\_adam, $s$\_adam\;
 $t \leftarrow$  $t^{th}$ iteration\;
 diff $\leftarrow$ zero\_matrix, size(diff) = size($p$)\;
 \For{i $\leftarrow$ 0, {\rm size(param)}}{
    \For{j $\leftarrow$ 0, {\rm size(param[$i$])}}{
        \For{k $\leftarrow$ 0, {\rm size(param[$i][j$])}}{
            $p$\_plus $\leftarrow$ copy($p$)\;
            $p$\_subs $\leftarrow$ copy($p$)\;
            $p$\_plus$[i][j][k]$ $\leftarrow p[i][j][k] + s$\;
            $p$\_subs$[i][j][k]$ $\leftarrow p[i][j][k] - s$\;
            diff$[i][j][k] \leftarrow$ QDP($p$\_plus, $p$\_subs, $c$, $s$)\;
}}}
$p$, $v$\_adam, $s$\_adam $\leftarrow$ Adam(diff, $p$, $v$\_adam, $s$\_adam, $\eta$, $\beta_{1}$, $\beta_{2}$, $\epsilon$, t) \;
 \caption{
 Adam optimization implementation in Qsun}
\label{al:GD}
\end{algorithm}

\subsubsection*{Preliminary QNN results}

\begin{figure*}[t!]
\centering
\includegraphics[width=14cm]{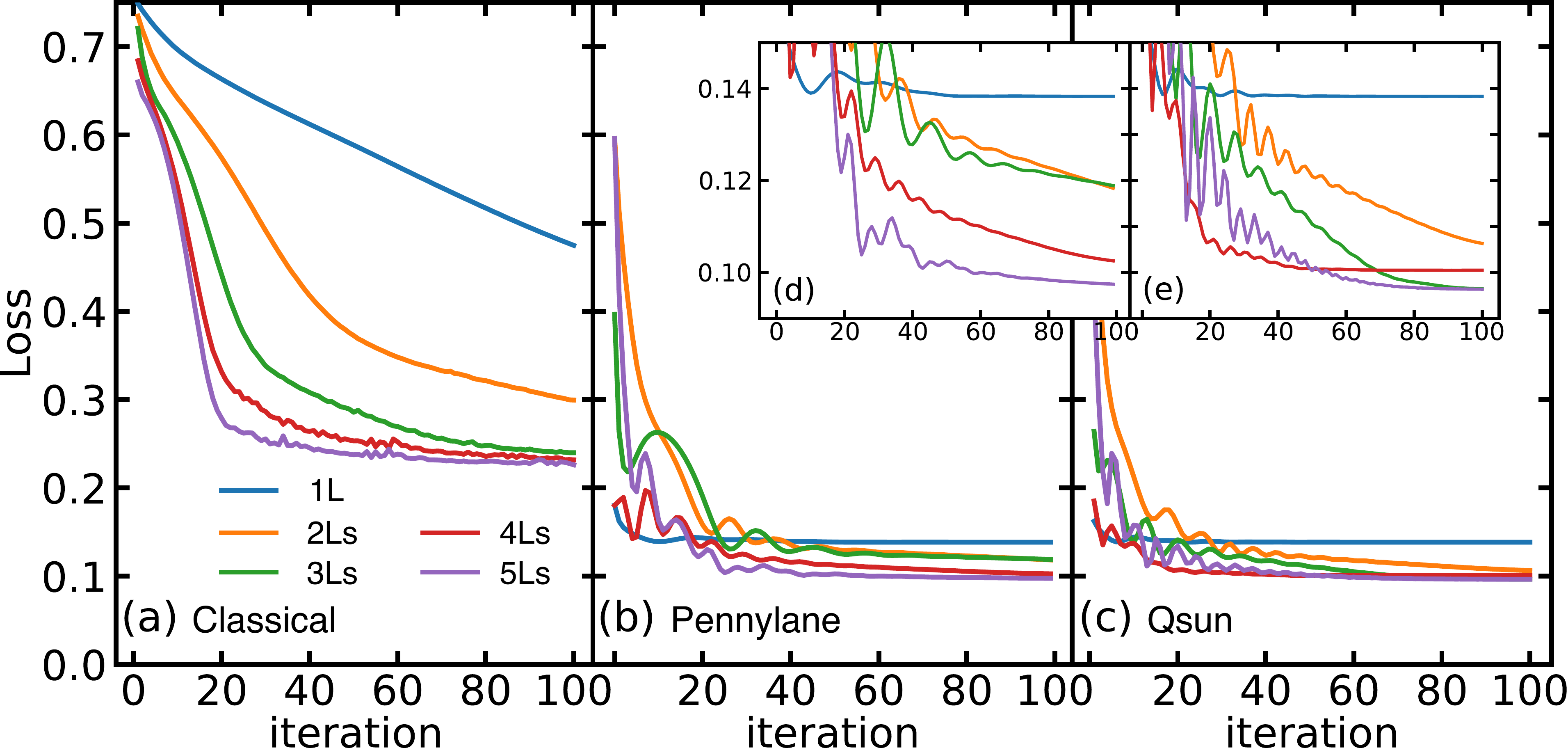} 
\caption{(a) Classical NN model; (b) and (c)
QNN models using Pennylane and
Qsun, respectively;
(d) and (e) are the 
zoom-in of (b) and (c), respectively. Each plot shows from 1 layer (1L)
to 5 layers (5Ls). See text for more details about parameters used for each model.}
\label{fig:compare_model}
\end{figure*}

We apply the QNN model on the Social Network Ads dataset  \cite{SNA}.
This is a categorical dataset to determine whether a user purchases a particular product or not.
We consider two features $Age$ and $EstimatedSalary$ 
to train the model with the output $Purchased$.
We use 400 samples and split them into 323 samples 
for the training set and 80 samples for the testing set. 
Once the data is normalized, 
we encode $Age$ and $EstimatedSalary$
into two qubits $|\psi_j\rangle, j = {0, 1}$, while $y$ = $Purchased$. 
We evaluate the predicted value $\widehat y$ and minimize the loss function (\ref{eq:cost2}) to train the QNN model.
The full code is given in \ref{appendix:qnn}.

For the comparison of performance, we train the QNN model using Qsun and Pennylane. We note that while the QDP and Adam algorithm are implemented in Qsun and Pennylane separately, the QNN circuit, encoding, and decoding procedures are the same for both approaches. We also compare their results with those from a classical model to explore advantages of QNN. For the classical NN model, we use the MLPClassifier function from scikit-learn that has 100 nodes per layer with the ReLU activation function. For the QNN model, we have 4 nodes per layer with the sigmoid activation function. We fix 
$s_0 = s_1 = \pi/20$ in the parameter-shift rule and $\eta = 0.1$.
For all optimization, we use
the Adam algorithm with $\beta_1 = 0.9,\beta_2 = 0.99$, and $\epsilon = 10^{-6}$.

Fig.~\ref{fig:compare_model} represents
the loss function versus iteration for different layers 
as shown in the figure for three cases: 
a classical NN model (a) and a QNN model trained by Pennylane (b) and Qsun (c).
In general, the loss function will reduce  and get unchanged after sufficient layers. 
In our example, the loss function becomes saturated after 5 layers. As expected, the loss function of QNN model
is much smaller than the the classical one.
Interestingly, when comparing the two quantum approaches (see insets (d, e) for a zoom in), the Qsun loss is converged faster than the Pennylane one.

\begin{figure*}[t]
\centering
\includegraphics[width=14.0cm]{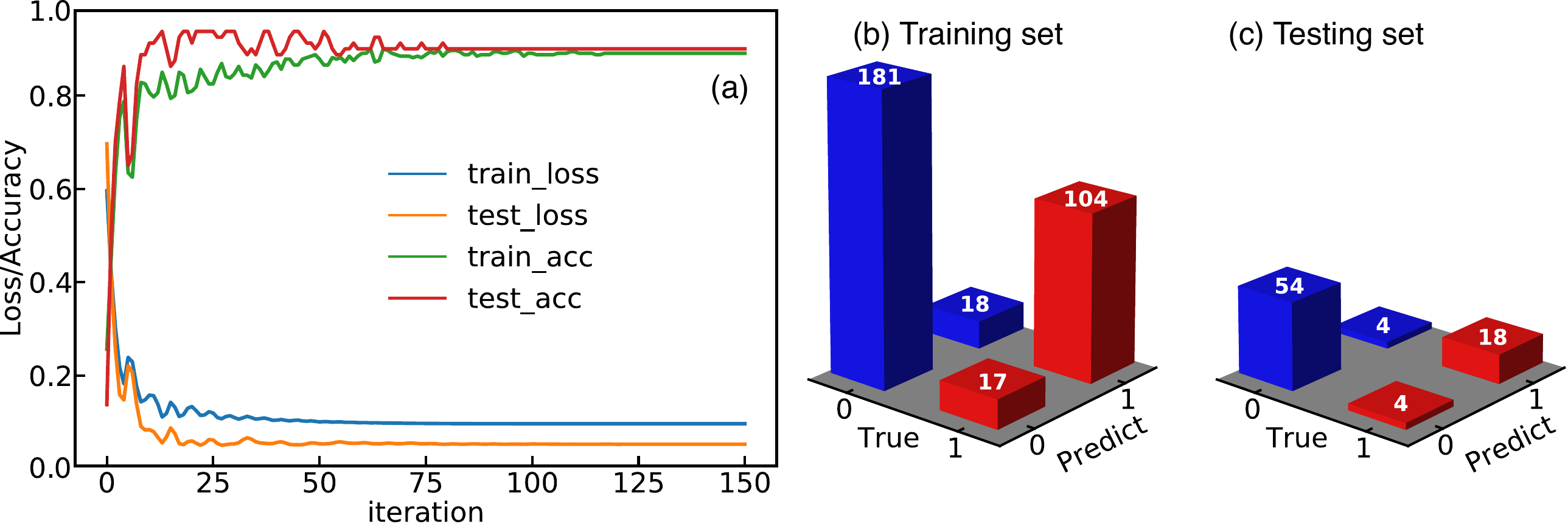}  
\caption{
Results from our QNN model with five layers 
and two features on the Social Network Ads dataset.
(a) Plot of loss and accuracy versus 
iteration for the training and testing process. 
(b, c) The confusion matrices for the training and testing sets.
Here we use 400 samples and split them 
into 320 samples for training 
and 80 samples for testing
with 400 iterations.}
\label{fig:qnn}
\end{figure*}

We next focus on our QNN model with fixed 5 layers.
In Fig.~\eqref{fig:qnn}(a),
we show the reduction of cost functions 
and activation functions during the iteration
for the training and testing processes. 
They both behave similarly and 
become saturated after 100 iterations.
Fig.~\eqref{fig:qnn}(b, c) 
represent results for the training and testing sets.
In the figure, both the true value $y$
and the predicted value $\widehat y$ 
are labeled as `0' and `1' 
for $no-purchase$ and $purchase$, respectively. 
$y\widehat y$ = `00' or `11' indicates a correct prediction,
whereas $y\widehat y$ = `01' or `10' 
indicates a wrong prediction.
It can be seen from Fig. \ref{fig:qnn} 
that our QNN model has a good performance, 
reflected by large values of 
main diagonal elements 
$y\widehat y$ = `00' and `11'. 
Importantly, unlike the classical 
neural network \cite{SNA}, 
it does not require many nodes in each layer, 
which is one of the main advantages of QNN. 

\section{Conclusion}\label{secv}
We have developed Qsun, an open-source platform of quantum virtual machine, that simulates the operation of a real quantum computer using the wavefunction approach. For small qubit numbers, the current version of Qsun runs standard tasks significantly faster than other platforms do. The quantum differentiable programming is implemented as a built-in function of Qsun, allowing us to execute quantum machine learning applications effectively. We have employed Qsun to implement two standard models in machine learning: linear regression and neural network. For the former, Qsun yields a quantum regression model nearly overlap with the classical reference, and somewhat better than the Pennylane one. For the latter, the QNN model trained using Qsun shows a good performance with a less number of nodes in each layer than the classical neural network. 
Although Qsun is aimed to quantum machine learning problems, as a generic quantum virtual machine, it can be used for multiple other purposes, such as developing and testing variational quantum algorithms for electronic structures or quantum information. It is well-fit to personal use thanks to its light weight. Qsun is under active development to cover a wide range of contents in machine learning. Our code is open source and available on GitHub \cite{QSUN}. 

\begin{acknowledgments}
This work was supported by JSPS KAKENHI Grant Number 20F20021.
\end{acknowledgments}

\bibliographystyle{unsrt}
\bibliography{references}


\appendix

\section{Code for benchmarking}
\label{appendix:test}
In this section, we present the code used to benchmark the performance of Qsun and compared it to other packages such as Qiskit, ProjectQ, and Pennylane. The criteria are the time to execute a certain quantum circuit of constant depth. The circuit contain an H, a $\sqrt{X}$, and a CNOT with \(depth = 10\) and measurements on all qubits. With the number of qubits as a variable, the time needed to execute the circuit is estimated, yielding the package's performance. The number of qubits varies from 2 to 15 qubits, which means we have to implement 14 circuits for each simulation. The implementation's time is stored in a Pandas's Dataframe and is shown in Fig. 2 in the main text.

\begin{lstlisting}[language=Python, caption={Implementation of the benchmarking test on four typical languages: Qsun, Qiskit, ProjectQ, and Pennylane.}]
#import module for visualization
import pandas as pd
import seaborn as sns
import time
import matplotlib.pyplot as plt

#import Qiskit module
from qiskit import IBMQ, BasicAer
from qiskit.providers.ibmq import least_busy
from qiskit import QuantumCircuit, ClassicalRegister, QuantumRegister, execute

#import ProjectQ module
from projectq import MainEngine
import projectq.ops as ops
from projectq.backends import Simulator

#import Qsun module
from Qsun.Qcircuit import *
from Qsun.Qgates import *

#import Pennylane module
import pennylane as qml
from pennylane import numpy as np

# Qsun circuit
def qvm_circuit(n_qubit, depth):
    circuit = Qubit(n_qubit)
    for m in range(depth):
        for i in range(n_qubit):
            H(circuit, i)
            Xsquare(circuit, i)
        for i in range(1, n_qubit):
            CNOT(circuit, i, 0)
    return circuit.probabilities()
    
# Pennylane circuit
dev = qml.device('default.qubit', wires=qubit_max)

@qml.qnode(dev)
def qml_circuit(n_qubit, depth):
    for m in range(depth):
        for i in range(n_qubit):
            qml.Hadamard(wires=i)
            qml.SX(wires=i)
        for i in range(1, n_qubit):
            qml.CNOT(wires=[i, 0])
    return qml.probs(wires=range(n_qubit))
    
# Qiskit circuit
def qiskit_circuit(n_qubit, depth):
    qr = QuantumRegister(n_qubit)
    cr = ClassicalRegister(n_qubit)
    circuit = QuantumCircuit(qr, cr)
    for m in range(depth):
        for i in range(n_qubit):
            circuit.h(qr[i])
            circuit.rx(math.pi/2, qr[i])
        for i in range(1, n_qubit):
            circuit.cx(qr[i], qr[0])
    circuit.measure_all()
    backend = BasicAer.get_backend('qasm_simulator')
    result = execute(circuit, backend=backend, shots=1024).result()
    return result.get_counts(circuit)
    
# ProjectQ circuit
def projectq_circuit(n_qubit, depth):
    eng = MainEngine(backend=Simulator(gate_fusion=True),engine_list=[])
    qbits = eng.allocate_qureg(n_qubit)
    for level in range(depth):
        for q in qbits:
            ops.H|q
            ops.SqrtX|q
            if q != qbits[0]:
                ops.CNOT|(q,qbits[0])
    for q in qbits:
        ops.Measure|q
    eng.flush()
    
#run test for Qsun
data_qvm = []

for n_qubit in range(qubit_min, qubit_max):
    start = time.time()
    qvm_circuit(n_qubit, depth)
    end = time.time()
    data_qvm.append(end-start)
    
#define the test parametes    
qubit_min = 2
qubit_max = 10
depth = 10    
    
#run test for Pennylane
data_qml = []

for n_qubit in range(qubit_min, qubit_max):
    start = time.time()
    qml_circuit(n_qubit, depth)
    end = time.time()
    data_qml.append(end-start)
    
#run test for Qiskit
data_qiskit = []

for n_qubit in range(qubit_min, qubit_max):
    start = time.time()
    qiskit_circuit(n_qubit, depth)
    end = time.time()
    data_qiskit.append(end-start)
    
#run test for ProjectQ
data_projectq = []

for n_qubit in range(qubit_min, qubit_max):
    start = time.time()
    projectq_circuit(n_qubit, depth)
    end = time.time()
    data_projectq.append(end-start)

df = pd.DataFrame({'QVM': data_qvm, 'Pennylane': data_qml, 
                   'Qiskit': data_qiskit, 'ProjectQ': data_projectq}, index = range(qubit_min, qubit_max))
                   
sns.lineplot(data = df)
plt.xlabel('Number of Qubits with depth = '+str(depth))
plt.ylabel('Time(s)')
plt.grid()
plt.savefig('compare.png')
\end{lstlisting}

\section{Code for QDP benchmarking}
\label{app_QDP}
For this code, we run the QDP algorithm on Qsun and measure how long it will take to update parameters. We implement the circuit described in Fig. \ref{fig:test_qdp} and measure the expected values of output, then record the time.
The number of qubits varies from 1 to 10, and the maximum number of iterations is 1000. Parameters for QDP, in this case, are \( \eta = 0.1 \) and \(s = \pi/20 \).

\begin{lstlisting}[language=Python, caption=Implementation of the quantum differentiable programming test in  Qsun.]
from Qsun.Qcircuit import *
from Qsun.Qgates import *
import numpy as np
import time

def circuit(params, n):
    c = Qubit(n)
    for j in range(0, 2*n, 2):
        RX(c, j//2, params[j])
        RY(c, j//2, params[j+1])
    return c

def output(params):
    c = circuit(params, len(params)//2)
    prob = c.probabilities()
    return -np.sum([i*prob[i] for i in range(len(prob))])
    
def cost(params):
    expval = output(params)
    return expval

def grad(params, shift, eta):
    for i in range(len(params)):
        params_1 = params.copy()
        params_2 = params.copy()
        params_1[i] += shift
        params_2[i] -= shift
        diff[i] = (cost(params_1)-cost(params_2))/(2*np.sin(shift))
    for i in range(len(params)):
        params[i] = params[i] - eta*diff[i]
    return params

time_qvm = []
n = 10
for i in range(1, n+1):

    start = time.time()    
    params = np.random.normal(size=(2*i,))
    diff = np.random.normal(size=(2*i,))

    for i in range(1000):
        params = grad(params, np.pi/20, eta=0.01)
    end = time.time() 
    
    time_qvm.append(end-start)
    print(cost(params))
\end{lstlisting}

\section{QDP Tutorial Appendix}
\label{appendix:tutorial}
This appendix demonstrates how to run a gradient descent algorithm using QDP in Qsun. We use \( \eta = 0.1 \) and \( s = \pi/20\) to find the derivative of a circuit. By doing that, we maximize the expectation values of one qubit. So the objective function we want to maximize is \(|\langle 1|\psi\rangle|^2 \), where \( \ket{\psi} \) is a quantum state of that qubit. 

\begin{lstlisting}[language=Python, caption=Implementation of the Gradient Descent algorithm by using QDP in Qsun.]
import numpy as np
from Qsun.Qcircuit import *
from Qsun.Qgates import *

def circuit(params):
    c = Qubit(1)
    RX(c, 0, params[0])
    RY(c, 0, params[1])
    return c

def output(params):
    c = circuit(params)
    prob = c.probabilities()
    return 0.*prob[0] + 1*prob[1]
    
def cost(params):
    c = circuit(params)
    prob = c.probabilities()
    expval = output(params)
    return np.abs(expval - 1)**2

def grad(params, shift, eta):
    for i in range(len(params)):
        params_1 = params.copy()
        params_2 = params.copy()
        params_1[i] += shift
        params_2[i] -= shift
        diff[i] = (cost(params_1)-cost(params_2))/(2*np.sin(shift))
    for i in range(len(params)):
        params[i] = params[i] - eta*diff[i]
    return params

params = np.random.normal(size=(2,))
diff = np.random.normal(size=(2,))

for i in range(1000):
    params = grad(params, np.pi/20, eta=0.01)
    
print('Circuit output:', output(params))
print('Final parameter:', params)

>>> Circuit output: -0.9381264201123851
>>> Final parameter: [ 0.29017649 -2.93657549]
\end{lstlisting}

\section{Quantum Linear Regression Appendix}
\label{appendix:regression}
Here, a Quantum Linear Regression model is programmed in Qsun, with results shown in Fig. \ref{fig:compare_linear_regression}. Its parameters are \( k = 10 \), \(\eta = 0.1\), and \(s = \pi/20 \). The optimization algorithm used in this code is Gradient Descent.

\begin{lstlisting}[language=Python, caption=Implementation of the linear regression model in qSUN.]
from Qsun.Qcircuit import *
from Qsun.Qgates import *
from sklearn import datasets

def circuit(params):
    c = Qubit(1)
    RX(c, 0, params[0])
    RY(c, 0, params[1])
    return c

def output(params):
    c = circuit(params)
    prob = c.probabilities()
    return 1*prob[0] - 1*prob[1]

def predict(x_true, coef_params, intercept_params, boundary=10):
    coef = boundary*output(coef_params)
    intercept = boundary*output(intercept_params)
    return coef*x_true.flatten() + intercept

def errors(x_true, y_true, coef_params, intercept_params, boundary):
    return mean_squared_error(y_true, predict(x_true, coef_params, intercept_params, boundary))

def grad(x_true, y_true, coef_params, intercept_params, shift, eta, boundary):
    
    coef_diff = np.zeros((2,))
    intercept_diff = np.zeros((2,))
    
    for i in range(len(coef_params)):
        coef_params_1 = coef_params.copy()
        coef_params_2 = coef_params.copy()
        coef_params_1[i] += shift
        coef_params_2[i] -= shift
        for x, y in zip(x_true, y_true):
            coef_diff[i] -= x*(y-predict(x, coef_params, intercept_params, boundary))*(output(coef_params_1)-output(coef_params_2))/(2*np.sin(shift))
            
        
    for i in range(len(coef_params)):
        intercept_params_1 = intercept_params.copy()
        intercept_params_2 = intercept_params.copy()
        intercept_params_1[i] += shift
        intercept_params_2[i] -= shift
        for x, y in zip(x_true, y_true):
            intercept_diff[i] -= (y-predict(x, coef_params, intercept_params, boundary))*(output(intercept_params_1)-output(intercept_params_2))/(2*np.sin(shift))
     
    coef_diff = coef_diff*boundary*2/len(y_true)
    intercept_diff = intercept_diff*boundary*2/len(y_true)
    
    for i in range(len(coef_params)):
        coef_params[i] = coef_params[i] - eta*coef_diff[i]
        
    for i in range(len(intercept_params)):
        intercept_params[i] = intercept_params[i] - eta*intercept_diff[i]
        
    return coef_params, intercept_params
    
X, y = datasets.load_diabetes(return_X_y=True)
y = (y - np.min(y)) / (np.max(y) - np.min(y))
# Use only one feature
X = X[:, np.newaxis, 2]

# Split the data into training/testing sets
X_train = X[:400]
X_test = X[-10:]

# Split the targets into training/testing sets
y_train = y[:400]
y_test = y[-10:]
    
coef_params = np.random.normal(size=(2,))
intercept_params = np.random.normal(size=(2,))

start = time.time()
for i in range(1000):
    coef_params, intercept_params = grad(X_train, y_train, coef_params, intercept_params, np.pi/20, eta=0.01, boundary=10)
end = time.time()

y_pred = predict(X_test, coef_params, intercept_params, boundary=10)
\end{lstlisting}

\begin{widetext}
\section{Quantum Neural Network Appendix}
\label{appendix:qnn}
The code showing in this appendix implements a process from training a QNN model with 5 layers and 2 qubits on the "Social\_Network\_Ads" dataset. From this dataset, we use a quantum neural network to predict whether a customer buys a car or not. The function train\_test\_split() splits the original dataset into the training and testing datasets. MinMaxScaler() scales the training dataset to [0, 1] before feeding it into the QNN circuit. The optimization algorithm in this model is the Adam optimization algorithm, as presented in the main text. The parameters used in the model are: $\rm test\_size = 0.2; \rm random\_state = 0; s = \pi/20; \eta = 0.1;\beta_{1} = 0.9, \beta_{2} = 0.99;  \epsilon = 10^{-6}; \rm number\_of\_iterations = 150$.

\begin{lstlisting}[language=Python, caption=Implementation of the QNN with two layers and two features in Qsun.]
# import libraries
from Qsun.Qcircuit import *
from Qsun.Qgates import *
from Qsun.Qmeas import *
from sklearn.model_selection import train_test_split
from sklearn.preprocessing import MinMaxScaler
from sklearn.metrics import confusion_matrix
import pandas as pd
import seaborn as sn
import matplotlib.pyplot as plt
import time

# one layer with full entanglement
def layer(circuit, params):
    circuit_layer = circuit
    n_qubit = len(params)
    for i in range(n_qubit):
        RX(circuit_layer, i, params[i][0])
        RY(circuit_layer, i, params[i][1])
    for i in range(n_qubit-1):
        CNOT(circuit_layer, i, i+1)
    CNOT(circuit_layer, n_qubit-1, 0)
    return circuit_layer

# encoding the features
def initial_state(sample):
    circuit_initial = Qubit(len(sample))
    ampli_vec = np.array([np.sqrt(sample[0]), np.sqrt(1-sample[0])])
    for i in range(1, len(sample)):
        ampli_vec = np.kron(ampli_vec, np.array([np.sqrt(sample[i]), np.sqrt(1-sample[i])]))
    circuit_initial.amplitude = ampli_vec
    return circuit_initial

# QNN circuit
def qnn(circuit, params):
    n_layer = len(params)
    circuit_qnn = circuit
    for i in range(n_layer):
        circuit_qnn = layer(circuit_qnn, params[i])
    return circuit_qnn

# QNN model
def qnn_model(sample, params):
    circuit_model = initial_state(sample)
    circuit_model = qnn(circuit_model, params)
    return circuit_model

# activation function
def sigmoid(x):
    return 1 / (1 + math.exp(-10*x))

# make a prediction
def predict(circuit):
    prob_0 = measure_one(circuit, 0)
    prob_1 = measure_one(circuit, 1)
    expval_0 = prob_0[1]
    expval_1 = prob_1[1]
    pred = sigmoid(expval_0-expval_1)
    return [pred, 1-pred]

# make a prediction for exp
def predict_ex(circuit):
    prob_0 = measure_one(circuit, 0)
    prob_1 = measure_one(circuit, 1)
    expval_0 = prob_0[1]
    expval_1 = prob_1[1]
    return [expval_0, expval_1]

# loss function    
def square_loss(labels, predictions):
    loss = 0
    for l, p in zip(labels, predictions):
        loss = loss + (1 - p[l]) ** 2
    loss = loss / len(labels)
    return loss

# loss function of QNN
def cost(params, features, labels):
    preds = [predict(qnn_model(x, params)) for x in features]
    return square_loss(labels, preds)
    
# https://d2l.ai/chapter_optimization/adam.html?highlight=adam
def adam(X_true, y_true, params, v, s, shift, eta, drop_rate, beta1, beta2, eps, iter_num):
    diff = np.zeros(params.shape)
    for i in range(len(params)):
        for j in range(len(params[i])):
            for k in range(len(params[i][j])):
                dropout = np.random.choice([0, 1], 1, p = [1 - drop_rate, drop_rate])[0]
                if dropout == 0:
                    params_1 = params.copy()
                    params_2 = params.copy()
                    params_1[i][j][k] += shift
                    params_2[i][j][k] -= shift
                    for x, y in zip(X_true, y_true):
                        circuit = qnn_model(x, params)
                        circuit_1 = qnn_model(x, params_1)
                        circuit_2 = qnn_model(x, params_2)
                        ex_plus = predict_ex(circuit_1)
                        ex_subs = predict_ex(circuit_2)
                        pred = predict(circuit)
                        diff_loss = ((-1)**y)*(-2)*(1-pred[y])*pred[y]*(1-pred[y])
                        diff_ex = 10*((ex_plus[0] - ex_subs[0]) - (ex_plus[1] - ex_subs[1]))/(2*np.sin(shift))
                        diff[i][j][k] += diff_loss*diff_ex
                                                
    diff /= len(y_true)
    v = beta1 * v + (1 - beta1) * diff
    s = beta2 * s + (1 - beta2) * (diff**2)
    v_bias_corr = v / (1 - beta1 ** (iter_num+1))
    s_bias_corr = s / (1 - beta2 ** (iter_num+1))
    params -= eta * v_bias_corr / (np.sqrt(s_bias_corr) + eps)
    
    return params, v, s

# source: https://www.kaggle.com/rakeshrau/social-network-ads
dataset = pd.read_csv('Social_Network_Ads.csv')
X = dataset.iloc[:, 2:-1].values
y = dataset.iloc[:, -1].values

# splitting dataset
X_train, X_test, y_train, y_test = train_test_split(X, y, test_size = 0.2, random_state = 0)

# scaling feature
scaler = MinMaxScaler()
X_train = scaler.fit_transform(X_train)
X_test = scaler.transform(X_test)

# create parameters  
n_layer = 5
params = np.random.normal(size=(n_layer, len(X_train[0]), 2,))
v = np.zeros(params.shape)
s = np.zeros(params.shape)

# training model
start = time.time()
for i in range(150):
    params, v, s = adam(X_train, y_train, params, v, s, 
                        shift=np.pi/20, eta=0.1, drop_rate=0.0, beta1=0.9, beta2=0.999, eps=1e-6, iter_num=i)

# confusion matrix
label = y_test
pred = [predict(qnn_model(x, params)) for x in X_test]
pred = np.argmax(pred, axis=1)
con = confusion_matrix(label,pred)
sn.heatmap(con, annot=True, cmap="Blues")
\end{lstlisting}
\end{widetext}

\end{document}